\documentclass{midl} % Include author names

\usepackage{mwe} % to get dummy images

\title[FCD Detection with TL and XAI]{Focal Cortical Dysplasia Type II Detection Using Cross Modality Transfer Learning and Grad-CAM in 3D-CNNs for MRI Analysis}

\midlauthor{\Name{Lorenzo Lasagni\nametag{$^{1,2}$}} \orcid{0000-0001-7693-3024} \Email{lorenzo.lasagni@unifi.it}\\
\addr $^{1}$ Department of Physics and Astronomy, University of Florence, Florence, Italy \\
\addr $^{2}$ Health Physics Department, Meyer Children’s Hospital IRCCS, Florence, Italy \AND
\Name{Antonio Ciccarone\nametag{$^{2}$}} \Email{antonio.ciccarone@meyer.it}\\
\Name{Renzo Guerrini\nametag{$^{3}$}} \Email{renzo.guerrini@meyer.it}\\
\addr $^{3}$ Neuroscience and Human Genetics, Meyer Children’s Hospital IRCCS, Florence, Italy \AND
\Name{Matteo Lenge\nametag{$^{3}$}} \Email{matteo.lenge@meyer.it}\AND
\Name{Ludovico D'Incerti\nametag{$^{3}$}} \Email{ludovico.dincerti@meyer.it}
}

\begin{document}

\maketitle

\begin{abstract}
Focal cortical dysplasia (FCD) type II is a major cause of drug-resistant epilepsy, often curable only by surgery. Despite its clinical importance, the diagnosis of FCD is very difficult in MRI because of subtle abnormalities, leading to misdiagnosis. This study investigates the use of 3D convolutional neural networks (3D-CNNs) for FCD detection, using a dataset of 170 subjects (85 FCD patients and 85 controls) composed of T1-weighted and FLAIR MRI scans. In particular, it investigates the benefits obtained from cross-modality transfer learning and explainable artificial intelligence (XAI) techniques, in particular Gradient-weighted Class Activation Mapping (Grad-CAM). ResNet architectures (ResNet-18, -34, and -50) were implemented, employing transfer learning strategies that used pre-trained weights from segmentation tasks. Results indicate that transfer learning significantly enhances classification accuracy (up to 80.3\%) and interpretability, as measured by a novel Heat-Score metric, which evaluates the model’s focus on clinically relevant regions. Improvements in the Heat-Score metric underscore the model’s seizure zone localization capabilities, bringing AI predictions and clinical insights closer together. These results highlight the importance of transfer learning, including cross-modality, and XAI in advancing AI-based medical diagnostics, especially for difficult-to-diagnose pathologies such as FCD.
\end{abstract}

\begin{keywords}
Deep Learning, Epilepsy, FCD, XAI, Gradcam, Transfer Learning, MRI
\end{keywords}

\section{Introduction}
Focal cortical dysplasia (FCD) is a developmental malformation of the cortical system that represents a major cause of drug-resistant focal epilepsy. Surgery remains one of the primary and most effective treatment options for patients with drug-resistant epilepsy. This make FCDs the most frequently surgically removed epileptogenic lesions in children \cite{guerrini2021focal} and rank as the third most common in adults \cite{blumcke2017histopathological}.
The diagnosis of FCD has been refined according to the latest classification of the International League Against Epilepsy (ILAE) classification (2022), which describes several characteristic imaging features. These include cortical thickening, blurring of the gray matter–white matter junction, gyration anomalies, and focal hyperintensities in the subcortical white matter that may extend to the ventricular system, forming the 'transmantle sign' \cite{widdess2006neuroimaging}. Despite these advancements, FCDs are often missed during routine neuroradiological evaluations, as the abnormalities can be subtle. Consequently, establishing an accurate diagnosis can be challenging, requiring extensive diagnostic procedures, including invasive electroencephalography (EEG). In particular, patients classified as 'MRI negative' — those with focal epilepsy but without discernible MRI abnormalities — have a reduced probability of undergoing epilepsy surgery and often experience poorer surgical outcomes \cite{bien2009characteristics, tellez2010surgical}.\\
Artificial intelligence (AI) models have been developed to help detect FCD \cite{spitzer2022interpretable}, using surface-based feature extraction, moreover a comparison between multiple existing algorithm and two nnU-Unet has been presented \cite{kersting2024detection}. However, achieving state-of-the-art results remains difficult in the absence of sufficiently large datasets for model training. This limitation is particularly impactful when individual centers attempt to train their own models on different tasks, such as segmentation versus classification, where they may have larger datasets available. The reuse of pre-trained models could offer a pathway to enhanced outcomes while preserving prior computational investments.\\
Transfer learning (TL) is widely used to enhance performance when working with limited datasets. A common approach involves utilizing pre-trained weights from ImageNet. However, recent studies have questioned the validity of using natural images for pre-training \cite{wen2021rethinking} and have demonstrated that within-domain transfer learning outperforms inter-domain pre-training \cite{heker2020joint}. The use of convolutional neural networks (CNNs) trained on segmentation tasks, either directly for classification \cite{jun2021medical, chen2024longformer, yang2024cmvim} or for feature extraction to be utilized by a secondary classifier \cite{tang2024radiation}, has demonstrated performance advantages. Another approach has been proposed, which concatenates segmentation and classification to improve classification results through multi-task transfer learning \cite{li2024segmentation}. However, based on our bibliographic research, the benefits of transfer learning (TL) from segmentation to classification have been only briefly explored in the context of Alzheimer's disease \cite{tam2024quantitative}.\\
Deep learning (DL) models hold promise in medical diagnostics, yet their adoption is often hindered by the 'black-box' nature of their decision-making processes \cite{watson2019clinical}. Clinicians need more than simple binary predictions; to increase the confidence in predictions, they seek transparent and interpretable insights, so as to also minimize biases induced by the dataset \cite{arrieta2020explainable, mahbooba2021explainable}. In response to these challenges, explainable artificial intelligence (XAI) has emerged as a crucial field, to address the need for clarity in predictions from machine learning models \cite{holzinger2019causability}.\\
Gradient-weighted Class Activation Mapping (Grad-CAM) which produces higher-impact pixel/voxel activation maps for model classification has emerged as an XAI technique of particular interest\cite{selvaraju2017grad}. This approach has demonstrated success in various medical imaging applications, including lung \cite{kumaran2024explainable}, breast \cite{liu2024breast}, and brain tumor detection, thereby improving model interpretability and leading to improved clinical confidence \cite{selvaraju2017grad}. Future works could investigate anyway a comparison with other XAI methods to evaluate quantitatively the attention to pathological regions such as Positive-gradient-weighted \cite{itoh2022positive}, shap \cite{lundberg2017unified} or for multiple pathology per scan with Grad-Cam++ \cite{chattopadhay2018grad}.\\
This paper explores how the use of heatmap-based interpretative techniques, in particular Grad-CAM, can bring advantages in FCD detection using neural networks. This is due to the possibility of having a greater transparency of the decision-making process offered by the highlighting of the image regions most impactful for the diagnostic reasoning of the model\cite{suara2023grad}. In fact, this study introduces 'Heat-Score', a new metric derived from heatmaps generated by Grad-CAM, to quantitatively evaluate the attention of the model on the most clinically relevant regions. By integrating the Heat-Score with traditional accuracy metrics, this research provides a comprehensive framework for assessing model performance, balancing classification efficacy with attention to pertinent anatomical features.\\
This investigation underscores the utility of cross-modality transfer learning and Grad-CAM in enhancing the interpretability of deep learning models for FCD detection and highlights the significance of the Heat-Score as a step toward the clinical validation and adoption of AI in medical diagnostics.

\section{Materials and methods}

\subsection{Dataset Description}
This study utilized an open pre-surgical MRI dataset comprising 85 individuals with epilepsy due to FCD type II and 85 age- and sex-matched healthy control participants. Of the 85 people with epilepsy who participated in the study, 35 (41.2\%) are female and 50 (58.8\%) are male with a mean age of 28.9 years. The dataset includes high-resolution isotropic 3D-T1 and 3D-FLAIR MRI sequences, as well as manually labeled regions of interest (ROIs) and associated clinical data. The imaging data were acquired on a single 3T MRI scanner at the Life and Brain Center (Bonn). Two distinct acquisition protocols were applied due to a scanner update in 2014, resulting in voxel sizes of either 1.0 $mm^3$ or 0.8 $mm^3$ for T1-weighted images. FLAIR images were isotropic with a voxel size of 1.0 $mm^3$. Before the update, an eight channel headcoil was used, after the update, a 32 channel headcoil was used. Further information can be found in the original paper \cite{schuch2023open}.

\subsection{Neural Network Architecture}
\begin{figure}[h!]
    \centering
    \includegraphics[width=0.99\textwidth]{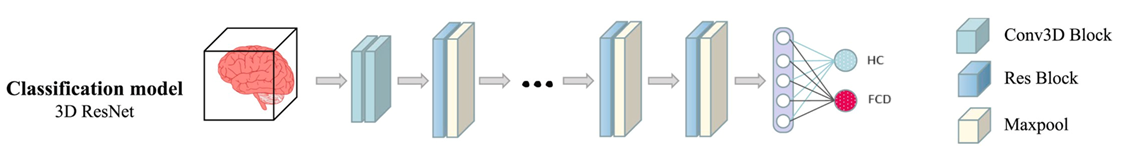} 
    \caption{Visual representation of 3D-ResNet used for our classification task.}
    \label{fig:resnet}
\end{figure}

A 3D convolutional neural network (3D-CNN) based on the 2D-ResNet \cite{targ2016resnet} architecture was employed for image classification. The network utilized residual blocks with two configurations: BasicBlock and Bottleneck, enabling varying depths and parameter complexities, a visual representation of the neural network is shown in Figure \ref{fig:resnet}. A ResNet-18, a ResNet-34 and a ResNet-50 were implemented for this study. The key steps implemented to adapt the original 2D-ResNet architecture into a 3D-ResNet model are summarized below.
\begin{itemize}
\item Input Layer: The network accepts single-channel 3D inputs and processes them with a 3D convolutional layer with a 7×7×7 kernel, a stride of 1×2×2, and batch normalization, followed by ReLU activation and max pooling.
\item Residual Blocks: The network includes four residual stages. Each block is designed to preserve the input features through identity mappings while applying 3D convolutions (3×3×3 kernels) and ReLU activation for feature extraction. Downsampling occurs in deeper blocks through strided convolutions.
\item Global Feature Aggregation: The features are aggregated using adaptive average pooling, reducing their spatial dimensions to a single 1×1×1 voxel per channel.
\item Classification Layer: A fully connected layer maps the 512-dimensional feature vector to a two-class classification output.
\end{itemize}

Our full-brain 3D classification model was trained on a workstation equipped with an NVIDIA GeForce RTX 4090 GPU (24 GB dedicated memory) and an Intel Core i9-13900K processor. The training time was approximately 2 to 3 hours per fold, with an inference time of $\sim$ 5 seconds per scan. While these hardware settings are non-standard, they remain accessible for adoption in smaller clinical centers.

\subsection{Training and Validation}
The 3D-ResNet models were trained using 5-fold cross-validation, where the dataset was divided into training and independent testing sets. The training subset was further partitioned to reserve a validation set, with an 80\%/20\% split. The following parameters were used during the training:
\begin{itemize}
\item Loss Function: Cross-entropy loss.
\item Optimizer: Adam optimizer with a starting learning rate of 0.001.
\item Scheduler: A ReduceLROnPlateau scheduler reduced the learning rate by a factor of 10 upon a validation loss plateau (patience: 5 epochs).
\item Regularization: Early stopping was applied based on validation performance to prevent overfitting.
\end{itemize}
Custom preprocessing steps included normalization and data augmentation. Input values were normalized using z-score normalization to achieve zero mean and unit variance. Data augmentation involved applying random rotations within a range of ±15 degrees to enhance model generalization.

\subsection{Transfer Learning Strategies}
Three distinct transfer learning strategies were evaluated, utilizing the weights provided by \cite{chen2019med3d}. These weights were derived from various ResNet models trained on one or 23 segmentation tasks. One segmentation task was based on a different medical imaging domain (lung CT scans), while the other 23 segmentation tasks may have been more beneficial, as they included one task on brain tumor segmentation, which is anatomically and modality-wise more closely related to FCD detection. In cases where transfer learning was applied, only the weights of the final block and layer could be modified.
\begin{itemize}
\item No Transfer Learning: Models were trained from scratch using random weight initialization.
\item Single Task Transfer Learning: Pre-trained weights derived from a single segmentation task were used to initialize the models.
\item Comprehensive Transfer Learning: Pre-trained weights derived from 23 segmentation tasks.
\end{itemize}

\subsection{Gradcam}
The explainability studies used a Layer GradCam method\cite{selvaraju2017grad}. Grad-CAM, to calculate the importance of different image regions in the classification, exploits the gradients of the model during the backpropagation phase. Then by mapping the gradients backwards, during the class activation, the importance of each region of the image for the final classification is obtained.\
This class activation is then superimposed on the original image to display the areas having the greatest impact on the final result. The Grad-CAM technique was selected for several reasons, including its ease of use and interpretability, which make it inherently accessible and easily understandable even for non-experts in neural networks. Furthermore, studies have demonstrated the utility of GradCam in aiding radiologists with diagnosis\cite{chien2022usefulness}, and it has been shown to outperform other proposed explainability techniques \cite{saporta2022benchmarking, zhang2021grad, lizzi2021convolutional}. Grad-CAM can make the neural network output more robust by providing a transparent and interpretable explanation. The ability to identify relevant regions allows potential biases or errors in classification to be identified\cite{selvaraju2017grad}. This can help highlight limitations or problems in the model, allowing targeted corrections or improvements to be made based on clinical information.\
To assess the quality assurance of the neural network, we introduced a novel parameter called Heat-Score, as far as our literature searches have revealed. This parameter enables the measurement of the correspondence between the regions of clinical interest for example, a potential epileptogenic region or any other suspicious pathological area, and the regions of interest utilized for determining the output of the neural network. The heatmap values generated by GradCam are normalized between 0 and 1 through a Min-Max normalization \cite{chattopadhay2018grad}. The Heat-Score is calculated as the average value of the heatmap voxels within the clinically relevant region of interest ($V_{S}$) minus the average value of the heatmap voxels within the background ($V_{Bkg}$), all divided by the standard deviation of the heatmap voxels in the background ($\sigma_{Bkg}$). This formula is analogous to the one used for calculating the Contrast-to-Noise Ratio (CNR). Opting to evaluate the average values between the regions eliminates the need to select an arbitrary threshold over the heatmap values for assessing the network's attention.

\begin{table}[h!]
\centering
\begin{tabular}{c c}
  % Left side table
  \begin{tabular}{c}
    $V_{S} = \text{Values of the Heatmap inside the Segmentation}$ \\
    $V_{Bkg} = \text{Values of the Heatmap in the Background}$ \\ 
    $\sigma_{Bkg} = \text{Standard Deviation of the Heatmap in the Background}$
  \end{tabular}
  &
  $HS = \frac{V_{S} - V_{Bkg}}{\sigma_{Bkg}}$  % Right side equation
\end{tabular}
\end{table}

Higher Heat-Score values indicate that in evaluating the test set, the neural network paid more attention to the same regions that a radiologist would have considered. In contrast, lower values indicate less understanding of the clinical problem and the possibility that attention was focused on artifacts or possible biases present in the dataset. Instead of performing a segmentation task, the use of the Heat-Score offers two key benefits. First, many real-world FCD cases lack segmentation masks, and even when available, there is no clear definition of the epileptogenic zone. A classification model that highlights suspicious regions can assist radiologists in identifying previously missed lesions and can be trained with imperfect segmentation, reducing the time required for annotation. Second, full segmentation models are computationally expensive and require larger datasets to achieve reliable results, as they perform voxel-wise classification. In contrast, classification models, even when trained on smaller datasets, can still contribute to the accurate diagnosis of MRI-negative cases.\\
Moreover, the Heat-Score introduces a valuable metric that can serve as a common language between software developers and clinicians. By quantifying how closely the model's attention aligns with expert judgment, it facilitates clearer communication about model performance, clinical relevance, and potential areas of improvement. This interpretability enables iterative development with more meaningful feedback.

\section{Results}
\begin{figure}[h!]
    \centering
    \includegraphics[width=0.7\textwidth]{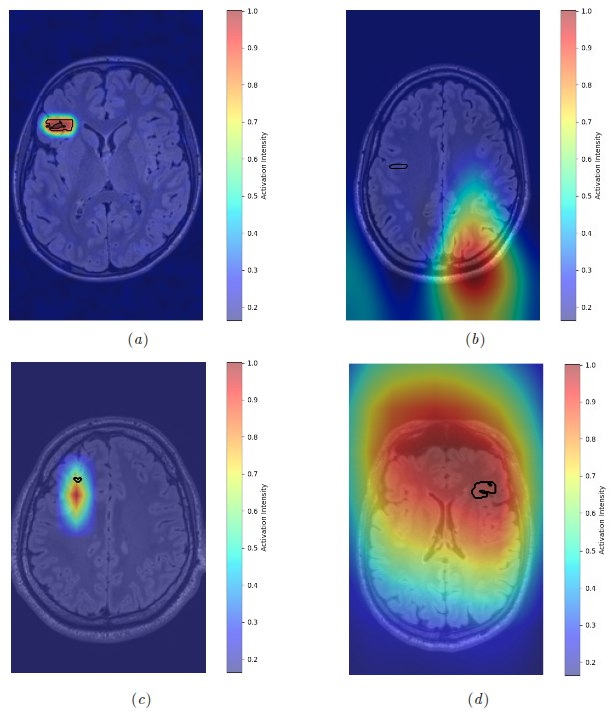} 
    \caption{An example illustrating a correct classification with accurate localization (a), a correct classification despite incorrect localization (b), a near-correct localization (c), and an overly vague localization (d) of FCD-II. The segmentation contour is outlined in black.}
    \label{fig:heatmap}
\end{figure}

Transfer learning from segmentation task consistently improved the classification accuracy across all ResNet models and MRI modalities and the results for T1w images can be found in Table \ref{Table:Results_T1} while for FLAIR images can be found in Table \ref{Table:Results_Flair}. For example, in the ResNet50 architecture, mean accuracy on FLAIR images increased from 72.9\% without transfer learning to 73.5\% with pretraining on 1 dataset and further to 80.3\% with pretraining on 23 datasets. A similar trend was observed in the ResNet18 and T1 modality, where the mean accuracy improved from 68.5\% (no transfer learning) to 76.6\% (1 dataset) and 80.3\% (23 datasets). This trend was consistent across other ResNet models, demonstrating the generalizability of transfer learning in enhancing the model's classification performance. The accuracy of our best model is higher than DeepFCD (Detection rate = 82\%, Specificity = 0\%), MELD (Detection rate = 49\%, Specificity = 55\%) and 3D-nnUNet (Detection rate = 55\%, Specificity = 86\%) compared to state-of-the-art results \cite{kersting2024detection}. However, it should be noted that our study was conducted on a single-center dataset, which may have contributed to higher performance results.\\
The Heat-Score, a novel metric reflecting the ability to localize the epileptogenic zone, also showed marked improvements with transfer learning. Without transfer learning, the Heat-Score for ResNet50 on FLAIR images was 1.545, increasing to 2.368 with pretraining on 1 dataset and peaking at 2.940 with pretraining on 23 datasets. For T1 images, the Heat-Score improved similarly from 2.297 (no transfer learning) to 2.815 (1 dataset) and 2.898 (23 datasets).\\
A paired t-test comparing models trained without transfer learning (TL) against those trained with 1 and 23 segmentation tasks, both in terms of accuracy and Heat-Score, was performed and found to be significant at $p < 0.05$. The comparison between TL with 1 segmentation task versus No TL yielded $p = 0.275$ for accuracy and $p = 0.042$ for Heat-Score. The comparison between TL with 23 segmentation tasks versus No TL resulted in $p = 0.007$ for accuracy and $p = 0.002$ for Heat-Score.\\ 
While both FLAIR and T1 modalities benefited from transfer learning, FLAIR images generally yielded higher Heat-Scores compared to T1 images, indicating better localization of the epileptogenic zone. This difference may reflect the greater sensitivity of FLAIR imaging to specific pathological features of FCD-II. Nevertheless, transfer learning reduced the performance gap between the modalities, with substantial Heat-Score improvements observed in both.\\
The standard deviations for accuracy and ROC-AUC were generally lower with transfer learning, especially when pretraining on 23 datasets. For instance, in ResNet50 on FLAIR, the standard deviation of accuracy decreased from 12.0\% (no transfer) to 7.0\% (23 datasets). This suggests that transfer learning not only improves performance metrics but also enhances the stability and reliability of the models. During the training phase, we observed a greater difference in accuracy on the validation set when no transfer learning (No-TL) was used across all MRI modalities, networks, and all five folds (No-TL: 0.061, 1-TL: 0.020, 23-TL: 0.015). This suggests increased overfitting and reduced generalizability in models trained from scratch compared to those benefiting from transfer learning.\\
Figure \ref{fig:heatmap} highlights the importance of assessing whether accurate classification is accompanied by a proper understanding of the underlying clinical abnormalities. It presents four examples of correct classification (presence of the FCD) with varying degrees of accuracy in localizing the pathological region: (a) correct identification of the affected area, (b) misidentification of the target region, (c) near-correct localization, and (d) overly vague localization. Cases (b), (c), and (d) received low Heat-Score values, while (a) received an high score. This distinction emphasizes the need to enhance the interpretability of neural networks to improve their reliability and trustworthiness among radiologists and neurologists.

\begin{table}[h!]
\centering
\begin{tabular}{|c|c|c|c|c|c|}
\hline
Model    & TL  & Modality & Accuracy ($\pm$ std) & ROC ($\pm$ std) & Heat-Score \\ \hline
resnet50 & no  & t1 & 0.685 ± 0.073 & 0.750 ± 0.090 & 2,297 \\ \hline
resnet50 & 1   & t1 & 0.766 ± 0.077 & 0.854 ± 0.054 & 2,815 \\ \hline
resnet50 & 23  & t1 & \textbf{0.791} ± 0.049 & \textbf{0.861} ± 0.063 & \textbf{2,898} \\ \hline
resnet34 & no  & t1 & 0.722 ± 0.044 & 0.780 ± 0.069 & 1,703 \\ \hline
resnet34 & 1   & t1 & 0.735 ± 0.051 & 0.849 ± 0.064 & 1,795 \\ \hline
resnet34 & 23  & t1 & 0.741 ± 0.038 & 0.832 ± 0.039 & 1,990 \\ \hline
resnet18 & no  & t1 & 0.698 ± 0.069 & 0.820 ± 0.061 & 1,118 \\ \hline
resnet18 & 1   & t1 & 0.741 ± 0.061 & 0.805 ± 0.078 & 1,634 \\ \hline
resnet18 & 23  & t1 & 0.784 ± 0.059 & 0.841 ± 0.057 & 2,618 \\ \hline
\end{tabular}
\caption{Performance results of different models with and without transfer learning for T1w images. The results refer to the test set.}
\label{Table:Results_T1}
\end{table}

\begin{table}[h!]
\centering
\begin{tabular}{|c|c|c|c|c|c|}
\hline
Model    & TL  & Modality & Accuracy ($\pm$ std) & ROC ($\pm$ std) & Heat-Score \\ \hline
resnet50 & no  & flair & 0.729 ± 0.120 & 0.771 ± 0.143 & 1.545 \\ \hline
resnet50 & 1   & flair & 0.735 ± 0.086 & 0.810 ± 0.148 & 2.368 \\ \hline
resnet50 & 23  & flair & \textbf{0.803} ± 0.070 & 0.817 ± 0.121 & \textbf{2.940} \\ \hline
resnet34 & no  & flair & 0.784 ± 0.086 & 0.819 ± 0.131 & 1.691 \\ \hline
resnet34 & 1   & flair & 0.778 ± 0.094 & 0.832 ± 0.094 & 2.117 \\ \hline
resnet34 & 23  & flair & 0.778 ± 0.086 & 0.821 ± 0.113 & 2.212 \\ \hline
resnet18 & no  & flair & 0.784 ± 0.104 & 0.862 ± 0.103 & 1.491 \\ \hline
resnet18 & 1   & flair & 0.760 ± 0.087 & 0.811 ± 0.081 & 1.870 \\ \hline
resnet18 & 23  & flair & \textbf{0.803} ± 0.062 & \textbf{0.864} ± 0.063 & 2.405 \\ \hline
\end{tabular}
\caption{Performance results of different models with and without transfer learning for FLAIR images. The results refer to the test set.}
\label{Table:Results_Flair}
\end{table}

\section{Conclusion}
The results of this study demonstrate that transfer learning, even when applied from a different task such as segmentation, significantly improves both the classification accuracy and the Heat-Score of ResNet models for detecting FCD-II in MRI images. These advancements were consistently observed across imaging modalities (FLAIR and T1) and ResNet architectures, highlighting the robustness of the approach. This work underscores the utility of transfer learning in addressing the challenges posed by limited data availability in the medical imaging domain and emphasizes its potential for cross-modality applications. Importantly, the enhancements in the Heat-Score metric suggest that transfer learning not only improves classification performance but also strengthens the model's ability to localize pathological regions with greater precision, as demonstrated by Grad-CAM heatmaps, thereby providing a valuable new metric for evaluating model interpretability and clinical relevance.\\
Despite these promising findings, some limitations must be acknowledged. The primary constraint of this study is the relatively small dataset size; however, to the best of our knowledge, this represents the largest publicly available dataset for FCD detection. Additionally, we recognize that training on a single-center dataset may limit the generalizability of our results to external test sets. To mitigate these concerns, we implemented a 5-fold cross-validation strategy and applied data augmentation techniques to enhance model robustness.
This study lays the groundwork for future works that should focus on validating these findings on larger, multicenter datasets and multi-FCD types to better assess the generalizability of the approach. Furthermore Heat-Score, while it is a novel metric to quantify how much a DL model focuses on pathologically ares, could be study better to verify how it behave with different explainable techniques.

\clearpage  % Acknowledgements, references, and appendix do not count toward the page limit (if any)
% Acknowledgments---Will not appear in anonymized version
\midlacknowledgments{We extend our gratitude to Leonardo Ubaldi for the insightful discussions on the application of machine learning. We also thank the MELD group for their unwavering support in fostering the development of deep learning expertise.}

\bibliography{main}

\begin{thebibliography}{33}
\providecommand{\natexlab}[1]{#1}
\providecommand{\url}[1]{\texttt{#1}}
\expandafter\ifx\csname urlstyle\endcsname\relax
  \providecommand{\doi}[1]{doi: #1}\else
  \providecommand{\doi}{doi: \begingroup \urlstyle{rm}\Url}\fi

\bibitem[Arrieta et~al.(2020)Arrieta, D{\'\i}az-Rodr{\'\i}guez, Del~Ser, Bennetot, Tabik, Barbado, Garc{\'\i}a, Gil-L{\'o}pez, Molina, Benjamins, et~al.]{arrieta2020explainable}
Alejandro~Barredo Arrieta, Natalia D{\'\i}az-Rodr{\'\i}guez, Javier Del~Ser, Adrien Bennetot, Siham Tabik, Alberto Barbado, Salvador Garc{\'\i}a, Sergio Gil-L{\'o}pez, Daniel Molina, Richard Benjamins, et~al.
\newblock Explainable artificial intelligence (xai): Concepts, taxonomies, opportunities and challenges toward responsible ai.
\newblock \emph{Information fusion}, 58:\penalty0 82--115, 2020.

\bibitem[Bien et~al.(2009)Bien, Szinay, Wagner, Clusmann, Becker, and Urbach]{bien2009characteristics}
Christian~G Bien, Miriam Szinay, Jan Wagner, Hans Clusmann, Albert~J Becker, and Horst Urbach.
\newblock Characteristics and surgical outcomes of patients with refractory magnetic resonance imaging--negative epilepsies.
\newblock \emph{Archives of neurology}, 66\penalty0 (12):\penalty0 1491--1499, 2009.

\bibitem[Blumcke et~al.(2017)Blumcke, Spreafico, Haaker, Coras, Kobow, Bien, Pf{\"a}fflin, Elger, Widman, Schramm, et~al.]{blumcke2017histopathological}
Ingmar Blumcke, Roberto Spreafico, Gerrit Haaker, Roland Coras, Katja Kobow, Christian~G Bien, Margarete Pf{\"a}fflin, Christian Elger, Guido Widman, Johannes Schramm, et~al.
\newblock Histopathological findings in brain tissue obtained during epilepsy surgery.
\newblock \emph{New England Journal of Medicine}, 377\penalty0 (17):\penalty0 1648--1656, 2017.

\bibitem[Chattopadhay et~al.(2018)Chattopadhay, Sarkar, Howlader, and Balasubramanian]{chattopadhay2018grad}
Aditya Chattopadhay, Anirban Sarkar, Prantik Howlader, and Vineeth~N Balasubramanian.
\newblock Grad-cam++: Generalized gradient-based visual explanations for deep convolutional networks.
\newblock In \emph{2018 IEEE winter conference on applications of computer vision (WACV)}, pages 839--847. IEEE, 2018.

\bibitem[Chen et~al.(2024)Chen, Fu, Bai, and Hong]{chen2024longformer}
Qiuhui Chen, Qiang Fu, Hao Bai, and Yi~Hong.
\newblock Longformer: longitudinal transformer for alzheimer's disease classification with structural mris.
\newblock In \emph{Proceedings of the IEEE/CVF winter conference on applications of computer vision}, pages 3575--3584, 2024.

\bibitem[Chen et~al.(2019)Chen, Ma, and Zheng]{chen2019med3d}
Sihong Chen, Kai Ma, and Yefeng Zheng.
\newblock Med3d: Transfer learning for 3d medical image analysis.
\newblock \emph{arXiv preprint arXiv:1904.00625}, 2019.

\bibitem[Chien et~al.(2022)Chien, Lee, Hu, and Wu]{chien2022usefulness}
Jong-Chih Chien, Jiann-Der Lee, Ching-Shu Hu, and Chieh-Tsai Wu.
\newblock The usefulness of gradient-weighted cam in assisting medical diagnoses.
\newblock \emph{Applied Sciences}, 12\penalty0 (15):\penalty0 7748, 2022.

\bibitem[Guerrini and Barba(2021)]{guerrini2021focal}
Renzo Guerrini and Carmen Barba.
\newblock Focal cortical dysplasia: an update on diagnosis and treatment.
\newblock \emph{Expert Review of Neurotherapeutics}, 21\penalty0 (11):\penalty0 1213--1224, 2021.

\bibitem[Heker and Greenspan(2020)]{heker2020joint}
Michal Heker and Hayit Greenspan.
\newblock Joint liver lesion segmentation and classification via transfer learning.
\newblock \emph{arXiv preprint arXiv:2004.12352}, 2020.

\bibitem[Holzinger et~al.(2019)Holzinger, Langs, Denk, Zatloukal, and M{\"u}ller]{holzinger2019causability}
Andreas Holzinger, Georg Langs, Helmut Denk, Kurt Zatloukal, and Heimo M{\"u}ller.
\newblock Causability and explainability of artificial intelligence in medicine.
\newblock \emph{Wiley Interdisciplinary Reviews: Data Mining and Knowledge Discovery}, 9\penalty0 (4):\penalty0 e1312, 2019.

\bibitem[Itoh et~al.(2022)Itoh, Misawa, Mori, Kudo, Oda, and Mori]{itoh2022positive}
Hayato Itoh, Masashi Misawa, Yuichi Mori, Shin-Ei Kudo, Masahiro Oda, and Kensaku Mori.
\newblock Positive-gradient-weighted object activation mapping: visual explanation of object detector towards precise colorectal-polyp localisation.
\newblock \emph{International Journal of Computer Assisted Radiology and Surgery}, 17\penalty0 (11):\penalty0 2051--2063, 2022.

\bibitem[Jun et~al.(2021)Jun, Jeong, Heo, and Suk]{jun2021medical}
Eunji Jun, Seungwoo Jeong, Da-Woon Heo, and Heung-Il Suk.
\newblock Medical transformer: Universal brain encoder for 3d mri analysis.
\newblock \emph{arXiv preprint arXiv:2104.13633}, 2021.

\bibitem[Kersting et~al.(2024)Kersting, Walger, Bauer, Gnatkovsky, Schuch, David, Neuhaus, Keil, Tietze, Rosenow, et~al.]{kersting2024detection}
Lennart~N Kersting, Lennart Walger, Tobias Bauer, Vadym Gnatkovsky, Fabiane Schuch, Bastian David, Elisabeth Neuhaus, Fee Keil, Anna Tietze, Felix Rosenow, et~al.
\newblock Detection of focal cortical dysplasia: Development and multicentric evaluation of artificial intelligence models.
\newblock \emph{Epilepsia}, 2024.

\bibitem[Kumaran~S et~al.(2024)Kumaran~S, Jeya, Khan, Alzahrani, and Alojail]{kumaran2024explainable}
Yogesh Kumaran~S, J~Jospin Jeya, Surbhi~Bhatia Khan, Saeed Alzahrani, and Mohammed Alojail.
\newblock Explainable lung cancer classification with ensemble transfer learning of vgg16, resnet50 and inceptionv3 using grad-cam.
\newblock \emph{BMC medical imaging}, 24\penalty0 (1):\penalty0 176, 2024.

\bibitem[Li et~al.(2024)Li, Wang, He, Chen, Wu, and Wu]{li2024segmentation}
Chen Li, Runyuan Wang, Ping He, Wei Chen, Wei Wu, and Yi~Wu.
\newblock Segmentation prompts classification: A nnunet-based 3d transfer learning framework with roi tokenization and cross-task attention for esophageal cancer t-stage diagnosis.
\newblock \emph{Expert Systems with Applications}, 258:\penalty0 125067, 2024.

\bibitem[Liu et~al.(2024)Liu, Himel, and Wang]{liu2024breast}
Suxing Liu, Galib Muhammad~Shahriar Himel, and Jiahao Wang.
\newblock Breast cancer classification with enhanced interpretability: Dalaresnet50 and dt grad-cam.
\newblock \emph{IEEE Access}, 2024.

\bibitem[Lizzi et~al.(2021)Lizzi, Scapicchio, Laruina, Retico, and Fantacci]{lizzi2021convolutional}
Francesca Lizzi, Camilla Scapicchio, Francesco Laruina, Alessandra Retico, and Maria~Evelina Fantacci.
\newblock Convolutional neural networks for breast density classification: performance and explanation insights.
\newblock \emph{Applied Sciences}, 12\penalty0 (1):\penalty0 148, 2021.

\bibitem[Lundberg and Lee(2017)]{lundberg2017unified}
Scott~M Lundberg and Su-In Lee.
\newblock A unified approach to interpreting model predictions.
\newblock \emph{Advances in neural information processing systems}, 30, 2017.

\bibitem[Mahbooba et~al.(2021)Mahbooba, Timilsina, Sahal, and Serrano]{mahbooba2021explainable}
Basim Mahbooba, Mohan Timilsina, Radhya Sahal, and Martin Serrano.
\newblock Explainable artificial intelligence (xai) to enhance trust management in intrusion detection systems using decision tree model.
\newblock \emph{Complexity}, 2021\penalty0 (1):\penalty0 6634811, 2021.

\bibitem[Saporta et~al.(2022)Saporta, Gui, Agrawal, Pareek, Truong, Nguyen, Ngo, Seekins, Blankenberg, Ng, et~al.]{saporta2022benchmarking}
Adriel Saporta, Xiaotong Gui, Ashwin Agrawal, Anuj Pareek, Steven~QH Truong, Chanh~DT Nguyen, Van-Doan Ngo, Jayne Seekins, Francis~G Blankenberg, Andrew~Y Ng, et~al.
\newblock Benchmarking saliency methods for chest x-ray interpretation.
\newblock \emph{Nature Machine Intelligence}, 4\penalty0 (10):\penalty0 867--878, 2022.

\bibitem[Schuch et~al.(2023)Schuch, Walger, Schmitz, David, Bauer, Harms, Fischbach, Schulte, Schidlowski, Reiter, et~al.]{schuch2023open}
Fabiane Schuch, Lennart Walger, Matthias Schmitz, Bastian David, Tobias Bauer, Antonia Harms, Laura Fischbach, Freya Schulte, Martin Schidlowski, Johannes Reiter, et~al.
\newblock An open presurgery mri dataset of people with epilepsy and focal cortical dysplasia type ii.
\newblock \emph{Scientific Data}, 10\penalty0 (1):\penalty0 475, 2023.

\bibitem[Selvaraju et~al.(2017)Selvaraju, Cogswell, Das, Vedantam, Parikh, and Batra]{selvaraju2017grad}
Ramprasaath~R Selvaraju, Michael Cogswell, Abhishek Das, Ramakrishna Vedantam, Devi Parikh, and Dhruv Batra.
\newblock Grad-cam: Visual explanations from deep networks via gradient-based localization.
\newblock In \emph{Proceedings of the IEEE international conference on computer vision}, pages 618--626, 2017.

\bibitem[Spitzer et~al.(2022)Spitzer, Ripart, Whitaker, D’Arco, Mankad, Chen, Napolitano, De~Palma, De~Benedictis, Foldes, et~al.]{spitzer2022interpretable}
Hannah Spitzer, Mathilde Ripart, Kirstie Whitaker, Felice D’Arco, Kshitij Mankad, Andrew~A Chen, Antonio Napolitano, Luca De~Palma, Alessandro De~Benedictis, Stephen Foldes, et~al.
\newblock Interpretable surface-based detection of focal cortical dysplasias: a multi-centre epilepsy lesion detection study.
\newblock \emph{Brain}, 145\penalty0 (11):\penalty0 3859--3871, 2022.

\bibitem[Suara et~al.(2023)Suara, Jha, Sinha, and Sekh]{suara2023grad}
Subhashis Suara, Aayush Jha, Pratik Sinha, and Arif~Ahmed Sekh.
\newblock Is grad-cam explainable in medical images?
\newblock In \emph{International Conference on Computer Vision and Image Processing}, pages 124--135. Springer, 2023.

\bibitem[Tam et~al.(2024)Tam, Liang, Chen, Wang, and Wu]{tam2024quantitative}
Thomas Yu~Chow Tam, Litian Liang, Ke~Chen, Haohan Wang, and Wei Wu.
\newblock A quantitative approach for evaluating disease focus and interpretability of deep learning models for alzheimer’s disease classification.
\newblock In \emph{2024 IEEE International Conference on Bioinformatics and Biomedicine (BIBM)}, pages 5104--5111. IEEE, 2024.

\bibitem[Tang et~al.(2024)Tang, Wang, Wu, Kong, Huang, and Han]{tang2024radiation}
Jingli Tang, Hao Wang, Dinghui Wu, Yan Kong, Jianfeng Huang, and Shuguang Han.
\newblock Radiation pneumonitis prediction using dual-modal data fusion based on med3d transfer network.
\newblock \emph{Journal of Imaging Informatics in Medicine}, pages 1--17, 2024.

\bibitem[Targ et~al.(2016)Targ, Almeida, and Lyman]{targ2016resnet}
Sasha Targ, Diogo Almeida, and Kevin Lyman.
\newblock Resnet in resnet: Generalizing residual architectures.
\newblock \emph{arXiv preprint arXiv:1603.08029}, 2016.

\bibitem[T{\'e}llez-Zenteno et~al.(2010)T{\'e}llez-Zenteno, Ronquillo, Moien-Afshari, and Wiebe]{tellez2010surgical}
Jos{\'e}~F T{\'e}llez-Zenteno, Lizbeth~Hern{\'a}ndez Ronquillo, Farzad Moien-Afshari, and Samuel Wiebe.
\newblock Surgical outcomes in lesional and non-lesional epilepsy: a systematic review and meta-analysis.
\newblock \emph{Epilepsy research}, 89\penalty0 (2-3):\penalty0 310--318, 2010.

\bibitem[Watson et~al.(2019)Watson, Krutzinna, Bruce, Griffiths, McInnes, Barnes, and Floridi]{watson2019clinical}
David~S Watson, Jenny Krutzinna, Ian~N Bruce, Christopher~EM Griffiths, Iain~B McInnes, Michael~R Barnes, and Luciano Floridi.
\newblock Clinical applications of machine learning algorithms: beyond the black box.
\newblock \emph{Bmj}, 364, 2019.

\bibitem[Wen et~al.(2021)Wen, Chen, Deng, and Zhou]{wen2021rethinking}
Yang Wen, Leiting Chen, Yu~Deng, and Chuan Zhou.
\newblock Rethinking pre-training on medical imaging.
\newblock \emph{Journal of Visual Communication and Image Representation}, 78:\penalty0 103145, 2021.

\bibitem[Widdess-Walsh et~al.(2006)Widdess-Walsh, Diehl, and Najm]{widdess2006neuroimaging}
Peter Widdess-Walsh, Beate Diehl, and Imad Najm.
\newblock Neuroimaging of focal cortical dysplasia.
\newblock \emph{Journal of Neuroimaging}, 16\penalty0 (3):\penalty0 185--196, 2006.

\bibitem[Yang et~al.(2024)Yang, Du, Yang, Du, Zheng, and Wang]{yang2024cmvim}
Guangqian Yang, Kangrui Du, Zhihan Yang, Ye~Du, Yongping Zheng, and Shujun Wang.
\newblock Cmvim: Contrastive masked vim autoencoder for 3d multi-modal representation learning for ad classification.
\newblock \emph{arXiv preprint arXiv:2403.16520}, 2024.

\bibitem[Zhang et~al.(2021)Zhang, Hong, McClement, Oladosu, Pridham, and Slaney]{zhang2021grad}
Yunyan Zhang, Daphne Hong, Daniel McClement, Olayinka Oladosu, Glen Pridham, and Garth Slaney.
\newblock Grad-cam helps interpret the deep learning models trained to classify multiple sclerosis types using clinical brain magnetic resonance imaging.
\newblock \emph{Journal of Neuroscience Methods}, 353:\penalty0 109098, 2021.

\end{thebibliography}

\end{document}